\documentclass[a4paper,preprint,aps]{revtex4}
\usepackage{graphicx}% Include figure files
\usepackage{dcolumn}% Align table columns on decimal point
\usepackage{bm}% bold math

\newcommand{\eqa}{\begin{equation}}
\newcommand{\eqz}{\end{equation}}
\newcommand{\eqma}{\begin{eqnarray}}
\newcommand{\eqmz}{\end{eqnarray}}

\begin{document}
%\Large
% title info
\title{Vibrational Spectra of the Azabenzenes Revisited: Anharmonic Force Fields}
\author{A. Daniel Boese}
\affiliation{Department of Organic Chemistry,
Weizmann Institute of Science, IL-76100 Re\d{h}ovot, Israel}
\author{Jan M. L. Martin}
\affiliation{Department of Organic Chemistry,
Weizmann Institute of Science, IL-76100 Re\d{h}ovot, Israel}
\date{{\it J. Phys. Chem. A} MS {\bf jp0369589}: Received October 2, 2003; Accepted November 14, 2003}
\smallskip
\begin{abstract}
Anharmonic force fields and vibrational spectra of the azabenzene series (pyridine, pyridazine,
pyrimidine, pyrazine, s-triazine, 1,2,3-triazine, 1,2,4-triazine and s-tetrazine) 
and benzene are obtained using  
density functional theory (DFT) with the B97-1 exchange-correlation functional and a triple-zeta plus
double polarization (TZ2P) basis set. Overall, the fundamental frequencies computed by second-order 
rovibrational perturbation
theory are in excellent agreement with experiment. The resolution of the presently calculated
anharmonic spectra is such that they represent an extremely useful tool for the assignment and
interpretation of the experimental spectra, especially where resonances are involved.
\indent
\end{abstract}
\maketitle  
\section{Introduction}

For atomization energies and geometries, density functional theory (DFT) constitutes a
cost-effective alternative to wavefunction-based {\it ab initio} methods, being capable of accuracies
on the order of a few kcal/mol and a few picometer, respectively, if basis sets of polarized triple
zeta quality are employed\cite{HCTH407,BMH}.

As a result of this astonishing success, especially considering the modest computational cost of DFT methods,
a variety of new applications are now being explored\cite{Colwell,Jamorski,Bauernschmitt,Carole,AmosDP},
including, very recently,  the calculation of molecular anharmonic potential energy surfaces \cite{thiel}.
Anharmonic potential energy surfaces (see Refs.\cite{Clabo88,Allen90} for initial pioneering studies at the
SCF and CISD level) allow for the prediction of molecular vibration-rotation spectra 
that can be compared directly to experiment, obviating the need to make estimations or approximations
for the effect of anharmonicity. CCSD(T)/$spdf$ anharmonic force 
fields can achieve accuracies on the order of 10 cm$^{-1}$ or better for fundamentals (e.g. \cite{c2h4} and 
references therein), and even greater accuracy is achievable if still larger basis sets and corrections for
inner-shell correlation are considered\cite{c2h2,diatomichydrides}.\\
DFT anharmonic force fields for small molecules have recently been the subject of 
two validation studies\cite{Neugebauer,AnharmVal}.
Our own validation study\cite{AnharmVal} suggested that, for fundamental frequencies, an RMS accuracy 
of about 18 cm$^{-1}$ can be attained when using
the B97-1 functional with sufficiently large basis sets. The question
arises as to how capable DFT is to treat medium-sized organic molecules in this fashion. 
Particularly relevant here is the study
the study of Miani et al.\cite{anharmhandy} on the fundamental frequencies of benzene, which
employed the B3LYP\cite{Bec93,Lee88} functional with the TZ2P (triple-zeta plus double polarization) basis set.
This paper, as well as a more recent study of furan, pyrrole and
thiophene\cite{Rudolf}, in fact yielded even more accurate results on these medium-sized organic
systems than suggested by the small-molecule validation studies cited above.\\
The azabenzenes are obtained by systematically replacing CH moieties in benzene by nitrogen atoms.
Numerous experimental spectra of these compounds are available in the literature, ranging from 
mono-azabenzene (pyridine) to 1,2,4,5-tetraazabenzene (s-tetrazine).  Pentazine is quite elusive\cite{Pentazide};
the equally elusive N$_6$ (hexazine
or hexaazabenzene) molecule has been the subject of extensive theoretical studies\cite{Saxe83,RJB2002}:
at the CCSD(T)/cc-pVTZ level, the hexazine (hexaazabenzene) structure undergoes distortion 
from the idealized $D_{6h}$ ring to a $D_2$ geometry,
which is a local minimum situated some 23 kcal/mol above the $C_{2h}$ diazide global minimum\cite{RJB2002}. 
Straka recently made the interesting suggestion that cyclic N$_6$ may form very stable M($\eta^6$-N$_6$) complexes
with M=Ti, Zr, Hf, Th\cite{Straka2002}.

In practice, nine molecules are relevant to the present study:  the parent benzene molecule, pyridine,
pyridazine, pyrimidine, pyrazine, 1,2,3-triazine, 1,2,4-triazine, 1,2,5-triazine ({\it sym}-triazine)
and 1,2,4,5-tetrazine (s-tetrazine). All molecules are displayed in figure 1.\\
Azabenzene skeletons serve as building blocks in nature, e.g., pyridine in pyridixol (vitamin B$_6$),
pyridazine and pyrimidine in pteridine, itself found in folic acid (vitamin vitamin B$_{10}$)
and riboflavin (vitamin B$_2$). Moreover the four DNA bases are derivates of pyrimidine (C, T)
and of purine (A, G) --- itself a pyrimidine-imidazole fused ring system. For these reasons,
azabenzene-like compounds figure in drug design, e.g. the reverse transcriptase inhibitors
AZT (contains pyrimidine) or nevirapine (contains pyridine).
Derivatives of pyridazine have been found to have potential therapeutic
or plant growth inhibitory effects.\\
Quite different applications include melamine-based plastics (melamine is a derivative of 
s-triazine) and the potential use of s-tetrazine in molecular data storage applications (one
of the relaxation mechanisms of the $S_1$ state leads to 2 HCN+N$_2$.\cite{Sch87,Luthi,tetr1,tetr2,tetr3}).\\
Innes, Ross and Moomaw (IRM throughout this paper) published a compilation and critical review of experimental
vibration spectroscopic data current to 1988\cite{Moomaw}. One of us (with C. Van Alsenoy)
carried out a harmonic-only B3LYP/cc-pVTZ study of the vibrational frequencies and geometries.\cite{first}
The limitations of this approach are obvious.\\

In the present contribution, we will consider DFT {\em anharmonic} force fields and anharmonic
vibrational spectra --- which are directly comparable to experiment --- for
the azabenzenes, and demonstrate their power as a spectroscopic assignment tool for medium-sized organic molecules.

\section{Computational Details}

In our validation study\cite{AnharmVal}, we considered a wide variety of exchange-correlation
functionals, as well as convergence in terms of the basis set and the integration grids
(both Kohn-Sham, KS, and Coupled Perturbed Kohn-Sham).

Especially for organic molecules, we found the best performance to be delivered by the B97-1
functional, which is Handy's reparametrization\cite{HCTH93} of Becke's 1997 hybrid functional\cite{B97}.
Satisfactory basis set convergence was generally found to be achieved with the TZ2P basis set\cite{TZ2P}.
This is therefore the functional/basis set combination employed for the present study.

We found\cite{AnharmVal} results for anharmonic force field calculations 
to be exceedingly dependent on the KS integration grid, much less so on the CPKS grid. For the
present study, we ended up using a (200,974) grid, that is, the direct product of a 200-point
Euler-MacLaurin radial grid with a 974-point Lebedev angular grid. For the CPKS steps,
we employed a (75,194) grid, which considerably reduced the overall computational cost.
Neither grid was pruned.

The quartic force fields were calculated by numerical central differences (in rectilinear
normal coordinates) of analytical
second derivatives, using a stand-alone driver program adapted from the CADPAC electronic structure
program system\cite{Cadpac}. The actual DFT second derivative calculations were done 
using the Gaussian 98 Rev. A11 program
package\cite{g98}. 

The step size in a numerical derivatives calculation always represents a compromise between
discretization error and roundoff error. Based on experience, we determined the step size
as a functions of both the absolute value of the harmonic frequency associated with the
normal coordinate involved, and the associated reduced mass
($\sqrt{\frac{1}{\sum_{j,k}(\frac{l_{MW,jk}(i)}{\sqrt{m_j}(i)})^2}}$):
\begin{eqnarray}
q_{step}(i)=4\times \sqrt{\frac{1}{\sum_{j,k}(\frac{l_{MW,jk}(i)}{\sqrt{m_j}(i)})^2}} \times
\sqrt{\frac{1000}{\omega(i)}}
\end{eqnarray}
In order to reduce roundoff error as much as possible, the KS and CPKS equations were basically
converged to machine precision.

To calculate this factor, both the reduced mass of the mode $m_j$ and the eigenvalues of the
normalisation matrix $l_{MW,jk}$ are needed.
The last part of equation 1, depending on the harmonic frequencies $\omega$, additionally
ensures that each displaced geometry has approximately the same energy difference to the minimum
geometry.\\
In this manner, we obtain a complete cubic force field, as well as all the diagonal and
semidiagonal quartic force constants. These are sufficient for second-order rovibrational
perturbation theory analyses\cite{rovib1,rovib2,rovib3,rovib4,rovib5}, which were carried out using the SPECTRO\cite{Spectro} and
POLYAD\cite{Polyad} programs developed in the Cambridge group and at Weizmann, respectively. 

\section{Results and Discussion}

\subsection{Benzene}

The parent molecule of our series, benzene (Figure 1a), has been the subject of a few previous anharmonic
force field studies, two at the SCF/DZP level\cite{MHAJ,HW}, and a very recent one at the
B3LYP/TZ2P level\cite{anharmhandy}. An extensive experimental literature exists on the subject
(see Refs.\cite{GOT,anharmhandy} for reviews): high resolution data are available for many of the
fundamentals, and all the assignments can be regarded as conclusive.
(An analysis of the B3LYP/cc-pVTZ 
normal modes of benzene and the azabenzenes
in terms of Pulay's redundant internal coordinates can be found in Table 17 of Ref.\cite{first}.) 

Experimentally obtaining harmonic frequencies for a molecule this large, even with such
high symmetry, is a nearly impossible task. In Table \ref{tab1} we compare computed harmonic frequencies with selected prior calculations
(in particular CCSD(T)/$spdf$ data\cite{benzccsdt}),
as well as several sets of experimentally derived data. 
The first such set are the `$\omega_{av}$'
estimates of  Handy, Murray and Amos (HMA)\cite{HMA}, themselves obtained as averages of an empirical 
estimate by Goodman, Ozkabak, and Thakur (GOT)\cite{GOT} and their own combination of experimental fundamentals with
the SCF/DZP computed anharmonicities of Maslen et al.\cite{MHAJ}. The second set are derived
from the experimental fundamentals and a new SCF/DZP anharmonic analysis by Handy and Willetts (HW)\cite{HW}.
The third set is derived from the experimental fundamentals and the B3LYP/TZ2P anharmonicities\cite{anharmhandy},
and is expected to be the most reliable.\\
Both sets of DFT numbers compare about equally well with the CCSD(T) data. The C-H stretching
frequencies are underestimated by about 20 cm$^{-1}$, while all other frequencies are being 
reproduced very accurately. (As an aside, we note that the C--H underestimation problem also
occurs\cite{benzeneMP2} at the M{\o}ller-Plesset perturbation theory (MP2) level.)
Although the C--H stretches are reproduced marginally worse by the B97-1 functional
(3 cm$^{-1}$) compared to the reference CCSD(T)/$spdf$ values, the overall performance is
improved: the mean absolute error decreases from 9.3 (B3LYP) to 6.5 (B97-1) cm$^{-1}$, and the
RMS deviation from 10.6 (B3LYP) to 9.6 (B97-1). The smaller decrease for the RMS error (compared to
the mean absolute error) can be
attributed to the greater proportional weight given to larger errors. When excluding the C-H stretching frequencies,
the RMS errors decrease to 8.4 (B3LYP) and 4.3 cm$^{-1}$ (B97-1).  Considering
the computational cost of density functional theory, the B3LYP results are
already of spectacular compared to the reference {\it ab initio} data. The B97-1 functional, however,
reduces the RMS error of the non C-H stretching frequencies by an additional 50\%.\\
Both functionals yield excellent agreement with experiment for the fundamental frequencies (Table \ref{tab2}).
The experimental data were taken from the earlier compilation of GOT\cite{GOT} and
from more recent gas-phase measurements\cite{anharmhandy}. The two sets of experimental data agree very
well with each other, except for $\nu_{16}$ and $\nu_{12}$ which are involved in strong Fermi resonances.
To obtain the B97-1/TZ2P fundamentals, we had to take the following Fermi resonances into
consideration: $\nu_3 + \nu_{16}$ and $\nu_{15}$ (with the deperturbed frequency $\nu_{15}$ at 3018
cm$^{-1}$ and the combination at 2947 $cm^{-1}$), $\nu_{13} + \nu_{16}$ and $\nu_{12}$
(unperturbed $\nu_{20}$ at 3040 and the combination band at 3082 $cm^{-1}$), $\nu_{16} + \nu_{13}$
and $\nu_5$ (with the unperturbed frequencies at 3064 and 3020 $cm^{-1}$), and $\nu_2 + \nu_{18}$
and $\nu_{16}$ (with $\nu_{16}$ (unperturbed) at 1600, $\nu_2 + \nu_{18}$ at 1603 $cm^{-1}$).\\
Interestingly, because of the small differences in the harmonic frequencies between the B3LYP/TZ2P
and the B97-1/TZ2P calculations and the different force field, now new Fermi resonances are
predicted. Furthermore, the fundamentals affected by these resonances are the ones which differ
the most when using the B3LYP or B97-1 functional.
Because of the Fermi resonances, the difference between different functionals and methods becomes
more important than on the harmonic level of approximation.
The results of both functionals compared to experiment are, however,
very similar, with mean absolute errors (compared to the latest experimental results of
Miani et al.\cite{anharmhandy}) of 9.4 (B3LYP)
and 8.3 (B97-1) cm$^{-1}$ and RMS errors of 17.0 and 13.7 cm$^{-1}$, respectively. Again,
all frequencies are very well described with the
exception of the C-H stretches, which are underestimated by 16--30
cm$^{-1}$ (discounting $\nu_5$ which is severely perturbed by resonances).
Without the C-H stretches, the mean absolute errors are reduced to 3.7 (B3LYP) and
3.9 (B97-1) cm$^{-1}$ and the RMS errors to 4.8 (B3LYP) and 4.5 cm$^{-1}$ (B97-1). Hence, while
clearly better for harmonic frequencies, the B97-1 functional is only slightly better for
predicting the fundamental frequencies of benzene. Nevertheless, the accuracy obtained by
density functional theory, which yields RMS errors of less than 5 cm$^{-1}$ for non C-H
stretching frequencies, is very good.\\
 In the third column, we added the B97-1/TZ2P anharmonic corrections to the CCSD(T)/ANO4321 
harmonic frequencies of Ref.\cite{benzccsdt}, in order to see how such a `hybrid' ab initio-DFT approach
would perform. The principal improvement seen is for the C--H stretches, which are thus all
bought into the range of the experimental values, except for $\nu_{5}$, which is in resonance
with $\nu_{16} + \nu_{13}$ as discussed above. When discounting C--H stretching frequencies, the
improvement compared to a pure B97-1/TZ2P calculation is quite modest, to 3.1
cm$^{-1}$ for the mean absolute error and to 4.1 cm$^{-1}$ for the RMS error. The corresponding
error statistics including C--H stretching frequencies change rather more significantly, 
to a mean absolute error of 4.0 cm$^{-1}$ and an RMS error of 7.4 cm$^{-1}$.\\
Summarizing, B97-1/TZ2P is likely to be a useful tool for analysis of vibrational spectra (and
verification of their assignments) of aromatic organic molecules in general and of the azabenzene
series in particular.

\subsection{Pyridine}

Pyridine being the simplest azabenzene (Figure 1b) and the closest to the parent molecule,
we may anticipate similar accuracy as for benzene.
As expected for this chemically important species, many experimental spectra are available,
in both liquid and gas phases. Most of the experimental results and the latest experimental
assignments are summarized by Klots\cite{Klots}. The computed and
observed vibrational frequencies are presented in Table \ref{tab3}.
As expected, the C-H stretching frequencies
are not well described. Here, they are also heavily perturbed, and, considering performance for
the corresponding bands in benzene, our method may not be sufficiently accurate to assist
in the experimental assignment. 
Thus, although we will attempt to elaborate on the C-H stretching frequencies for most azabenzenes,
the results are more tentative.
Additionally, a small change in
the original unperturbed fundamental frequencies will cause large changes in the perturbed
frequencies. Nevertheless, the two asymmetric stretches $\nu_{14}$ and $\nu_{15}$ do not follow
the general trend of being underestimated by 20--40 cm$^{-1}$. While the perturbed
$\nu_{14}$ is close to the experimental value, $\nu_{15}$ differs by 70 cm$^{-1}$.
Especially the latter appears too large to be accounted for merely by deficiencies in our
DFT calculation. Here, the severe Fermi resonance seems to be responsible for this
assignment, as its deperturbed value is at 3021 cm$^{-1}$. This mode is resonating with
(deperturbed frequencies given)
$\nu_{14}$ at 3053 cm$^{-1}$, $\nu_4 + \nu_{17}$ at 3015 cm$^{-1}$ and $\nu_{16} + \nu_5$ at 3054
cm$^{-1}$, with the perturbed $\nu_4 + \nu_{17}$ now appearing at 3023 cm$^{-1}$. It appears that
the latter value is more in line with experiment, and that perhaps this assignment might be more
reasonable. Nevertheless, some caution is appropriate as the C--H harmonic stretching frequencies are
not as well described by DFT as the remaining modes. Large basis set CCSD(T) harmonic frequencies
would be helpful here, but would require inordinate amounts of CPU time.\\
 Another picture emerges for the non C-H frequency modes. Here, we primarily compare to the
medium resolution Raman vapor data of Klots (Klots2)\cite{Klots}: their assignments seem to be in
line with most other experiments, although they differ from the most recent data obtained by Partal et
al.\cite{Kearley1}. Comparing to these latter inelastic neutron scattering (INS) data,
however, amounts to comparing apples and oranges, as the INS data are based on further
refinement of a DFT computed {\em harmonic} force field by maximizing agreement between simulated
and observed INS spectral. The pseudo-harmonic frequencies 
from the refined force field correspond neither to true harmonic nor to true 
fundamental frequencies --- in effect, they are neither fish nor fowl.\\
We can confirm the assignment of Klots\cite{Klots} of $\nu_{21}$ to the lower value of 1053
cm$^{-1}$. Difficult assignments seem to result from the Fermi resonances for both
$\nu_4$ and $\nu_{19}$, since these are the only fundamentals which give errors of 10
cm$^{-1}$ or larger. The deperturbed mode of $\nu_4$ at 1578 cm$^{-1}$ resonates
with $\nu_9 + \nu_{10}$ at 1593 cm$^{-1}$ resulting in perturbed modes at 1575 and
1596 cm$^{-1}$. Klots may have misassigned the latter band to $\nu_4$ rather than $\nu_9 + \nu_{10}$.
$\nu_{19}$, on the other hand, seems to be very difficult to
assign from experimental data because of its depolarized character. The problem is further
exacerbated 
as $\nu_{19}$ is located at the shoulder of $\nu_6$. Although $\nu_{19}$ is involved in a Fermi
resonance, it is probably not the source of the disagreement with, since its unperturbed
frequency at 1241 cm$^{-1}$ resonates with $\nu_{10} + \nu_{12}$ at 1282 cm$^{-1}$, resulting in
bands at 1237 and 1289 cm$^{-1}$. With the exception of these two strongly perturbed bands ($\nu_4$ and
$\nu_{19}$), the agreement between experiment and theory can only be described as stunning for
bands other than C-H stretches. For the latter, probably only a full CCSD(T) force field can resolve the
assignment. 
Even when including $\nu_4$ and $\nu_{19}$
(but excluding all C-H stretching frequencies), the mean absolute error for the computed
values  compared to the experimental data of Klots\cite{Klots} is 3.6 cm$^{-1}$ and
the RMS error is 5.3 cm$^{-1}$; without $\nu_4$ and $\nu_{19}$ they are reduced to 2.7 and
3.8 cm$^{-1}$, respectively.\\
The double-harmonic infrared intensities are in good agreement with the B3LYP/cc-pVTZ values\cite{first} and
in reasonable agreement with the experimental numbers. Even here, the C-H stretches differ,
probably not only because of anharmonic contributions and Fermi resonances, but also because
of the poor description of these modes by DFT. Still, the excellent performance seen for 
benzene results is repeated for pyridine.

\subsection{Pyridazine}

Pyridazine (Figure 1c)  has been of particular interest in the last couple of years
\cite{Kearley2,Holly,Boggs}. A large body of experimental data is available, generally measured in the
gas phase but when certain modes were unavailable, the authors of Ref.\cite{Holly,Boggs} substituted
their own liquid or solid phase measurements in order to get a full complement of frequencies.
All of the spectra have been assigned with the aid of scaled Hartree-Fock, MP2, 
BLYP or B3LYP harmonic force fields using small basis sets. 
Some of them have their
force fields fitted to match the experimental data. Despite the large amount of data, huge
discrepancies exist, as shown in Table \ref{tab4}. This seems to be especially true for the out-of-plane 
modes where 
experimental assignments and values seem almost arbitrary. 
For $\nu_{11}$, for example,
the experimental numbers range from 765 to 949 cm$^{-1}$, and even the two most recent numbers appear at 
opposite ends of this range. 
These assignments have been done using different
methods and scaling techniques, since no reliable theoretical data were available at the time.\\
The same problem with the C-H stretches that was encountered for benzene and pyridine occurs for
pyridazine as well; the situation is further complicated since the experimental data differs and thus
we cannot assign the frequencies of the experiment. Despite the discrepancies between various
experimental numbers, all the C-H stretches appear plausible within the error range of our method.\\
 As the INS data set dramatically varies from the remainder of the experimental data, and because it did
not appear to be reliable for pyridine, it was excluded from our analysis, although it
is reported in Table \ref{tab4} for completeness.\\
 Since such a large amount of experimental data is available, it is best to discuss each mode
separately, starting with the in-plane modes.
\begin{itemize}
\item For $\nu_3$, all experiments except the first of Ref.\cite{Ozono} agree on an assignment to
the 1570 cm$^{-1}$ band.
 Our results, however, would seem to indicate that the assignment around
1555 cm$^{-1}$ might be more plausible. 
\item The same occurs for $\nu_4$. Experimental values of Refs.\cite{Ozono,Boggs} cluster around 1440 cm$^{-1}$, with
Refs. \cite{Holly,Tucci} around 1415 cm$^{-1}$. Our calculation clearly favors the former assignment.
\item For $\nu_5$, significant differences between experiments of Refs.\cite{Ozono,Holly,Tucci}
and the computed
value of 1145 cm$^{-1}$ can be found. The experimental results are thus untenable in light of this huge
discrepancy of more than a hundred cm$^{-1}$, and only Ref.\cite{Tucci} proposes an assignment of 1160
cm$^{-1}$, which still appears to be on the high side.
\item For $\nu_6$, most experiments have assignments around the computed 1157 cm$^{-1}$;
however, Ref.\cite{Boggs} assigned this mode to 1120 cm$^{-1}$ because of their previous assignment.
Our calculated $\nu_5$ and $\nu_6$ nearly coincide,
\item For $\nu_7$, all experimental datasets are in agreement with each other and our calculations.
Considering the generally small anharmonicities for this type of vibrations, the discrepancy of 40 cm$^{-1}$ between the
computed $\omega_8$ and the observed $\nu_8$ seems to be a bit on the large side.
The experimental assignment may have been complicated by the band's position 
in an IR band envelope going from 960 to 980 cm$^{-1}$\cite{Boggs}.
\item As with $\nu_5$, the experimental assignment for $\nu_9$ of Vazquez et al.\cite{Boggs} differs from
the other obtained fundamentals, and is the only one that can be confirmed by our calculations.
\item For $\nu_{16}$, only Refs.\cite{Ozono,Holly} differ from the computed values, while for $\nu_{17}$,
the experimental assignment of Stidham and Tucci\cite{Tucci} both in the IR and the Raman phase appear
doubtful. $\nu_{18}$ agrees nicely with all experiments.
\item For $\nu_{19}$ and $\nu_{20}$, the same assignment problems that we experienced for $\nu_5$ and $\nu_6$
seem to have occurred; while most experiments assign values above 1100 cm$^{-1}$ to $\nu_{19}$,
only Vazquez et al.\cite{Boggs} seem to give the correct assignment. This results, however, in the
value of $\nu_{20}$ thus ending up a bit on the low side, and only the value of Ref.\cite{Tucci} ends up
close to our calculated number. The same seems to have happened to $\nu_{21}$, where only
Ref.\cite{Boggs} is in line with our calculations. Hence, for the
in-plane modes, the data set of Vazquez et al.\cite{Boggs} appears to be most
reliable, excluding $\nu_3$ and $\nu_6$ (which appear to be dubious assignments) and perhaps $\nu_8$ and $\nu_{20}$.
\item For the out-of-plane modes, the experimental assignment appears to be even more problematic.
For $\nu_{22}$, all experiments differ by 20 cm$^{-1}$, and are either found at 998 or 842 cm$^{-1}$.
Both assignments appear to be unlikely, since the harmonic B97-1 frequency is at 981 cm$^{-1}$, which would
put the fundamental too low for 998 and definitely way too high for assignment to the 842 cm$^{-1}$ band.
\item For $\nu_{23}$, all experimental assignments are around the computed 751 cm$^{-1}$, while for $\nu_{24}$,
all the experiments appear to yield assignments higher than our computed fundamental of 362 cm$^{-1}$.
\item Based on our data, all the $a_2$ symmetry modes ($\nu_{10}$ through $\nu_{13}$) should probably be re-examined, since
several experimental results differ drastically from each other and from our calculations. A
discussion of these modes is almost impossible, and with the exception of 
Refs. \cite{Boggs,Ozono} for $\nu_{13}$, all experimental assignments lie beyond the likely error bars
of the theoretical values.
\end{itemize}
On the whole, the pyridazine vibrational spectra definitely merit further investigation,
as the many experimental and theoretical datasets available are considerably at variance with each other.
Thus, it is impossible to give an error estimate for the method used based on
this molecule. However if our computational results prove to be as accurate as for benzene
and pyridine, the experimental spectra have to be reexamined. Based on this study, one
might wonder whether the addition of more nitrogens or the N-N bond is causing the deterioration of
the calculated values. 

\subsection{Pyrimidine}

Pyrimidine (Figure 1d), has been studied less in recent years. Notable are the IR experiments
compiled by IRM\cite{Moomaw}, the IR experiments of Billes et
al.\cite{Holly}, and the inelastic neutron scattering work of Navarro
et al.\cite{Kearley3}. Here, we compare mainly to the IR experiments, since the INS
data for neither pyridine nor pyridazine agree well with our computed values and
the other experiments, for reasons outlined above. 
Nevertheless, for pyrimidine the INS data
seem a lot closer to the IR
data than for the previous two azabenzenes (Table \ref{tab5}). The values for the $a_1$ C-H
stretches are in agreement with experiment, although this seems fortuitous. Otherwise, all
in-plane modes are within 8 cm$^{-1}$ of experiment, with the sole exception of $\nu_{17}$.
Interestingly, both $\nu_7$ and $\nu_9$ undergo strong Fermi type 1 and 2 resonances; 
their resonance matrices are displayed in Table\ref{tab6} with their respective eigensolutions.
For $\nu_7$, very small changes in the unperturbed fundamentals would result
in the value ending up not at 1071 cm$^{-1}$, but at the experimental assignment of 1065 cm$^{-1}$.
The assignment of the two resonating modes appears to be purely arbitrary. As for $\nu_9$,
the resonating overtone of $\nu_{24}$ matches the experimental number exactly. Since
both modes are in resonance, the overtone might have ``borrowed intensity''
(as described in Ref.\cite{Handyresonance}) from $\nu_9$, and thus the assignment.\\
 When looking at the out-of-plane modes in Table \ref{tab5}, only $\nu_{10}$ and $\nu_{20}$
yield different results compared to the first set of experimental results.
Both discrepancies have been mentioned before by the harmonic-only study\cite{first}, and
$\nu_{20}$ has a very small intensity. Moreover, by using scaled HF frequencies,
Wiberg\cite{Wiberg} has suggested that the assignment for $\nu_{10}$ might also be incorrect.
The IRM\cite{Moomaw} assignments
for modes 5, 6, 13 and 21 appear more plausible than from
Ref.\cite{Holly}. Neglecting the problematic modes 10 and 20 and reassigning mode 9 to 677 cm$^{-1}$,
the errors, including the C-H stretches, are 6.3 (mean absolute) and 11.4 cm$^{-1}$ (RMS),
and without them are 4.1 and 5.3 cm$^{-1}$, respectively. Thus this method again appears to
be very reliable
for predicting fundamental frequencies.

\subsection{Pyrazine}

As with pyrimidine, only few experiments have been done for pyrazine (Figure 1e).
The vibrational assignments are displayed in Table \ref{tab7}. In Ref.\cite{Holly}, assignments
were done in the gas phase for the IR active modes and in a melt for the Raman spectrum.
Overall, the agreement is again very good for modes other than C-H stretches. 
If the previously 
mentioned tendency to  underestimate the C-H stretches is taken into account, all assignments
appear plausible, with the possible exception of
the assignment of $\nu_{11}$ from Ref.\cite{Holly}.
It should be noted that all of the C-H stretches are perturbed by Fermi resonances.
For the in-plane modes, especially $\nu_{21}$ appears to lie beyond the error bars of our method. 
It has been mentioned by the harmonic-only study in Ref.\cite{first} that this `Kekul\'e mode'
is problematic and it was concluded that this mode has significant multireference character,
since it corresponds to the dissociation into 2 HCN + C$_2$H$_2$. However, hybrid density
functionals are usually capable to describe nondynamical correlation to a
certain extent, since for example the atomization energy of ozone is reasonably well reproduced.
For this mode, the assignment of Ref.\cite{Marcos} appears to be out of question.\\
For all the other modes, most the experiments scatter only by about 5 cm$^{-1}$, with this
agreement suggesting that their assignments and values are pretty accurate.
The out-of-plane modes are generally quite well described by theory. Note that the two $a_u$ modes
since they are neither IR nor Raman active; for this reason they are not assigned in
Ref.\cite{Holly}. The INS values are somewhat disappointing for both frequencies,
since the chief advantage of this method is the ability to predict such bands. Another discrepancy
is the first $b_{2g}$ mode ($\nu_9$), although the second experiment of Ref.\cite{Holly} is closer to our values.\\
This
leads to mean absolute and RMS errors compared to Ref.\cite{Holly} (Ref.\cite{Marcos}) of 7.8 (7.9) and
10.4 (11.1) cm$^{-1}$ including the C-H stretches, and of 5.4 (5.1) and 6.9 (6.5) cm$^{-1}$
when excluding them. If the problematic $\nu_2$ and $\nu_9$ are excluded, the mean absolute
and RMS errors for the non C-H stretches reduces to 4.1 (4.1) and 5.4 (5.5) cm$^{-1}$,
respectively. Thus, the general RMS discrepancy of 5 to 7 cm$^{-1}$ is obtained for
pyrazine as well.\\

\subsection{1,3,5 Triazine (s-Triazine)}

1,3,5-Triazine (Figure 1f) is the only azabenzene with a degenerate point group symmetry ($D_{3h}$).
Because of this, the $a_2'$ modes ($\nu_4$ and $\nu_5$) are both IR and Raman inactive. The
INS \cite{Kearley5} value for $\_nu_4$ coincides exactly with our calculation (Table \ref{tab8});
the discrepancy of 124 cm$^{-1}$ for $\nu_5$ is outlandish even for the INS data and this band is probably
misassigned.
The inelastic neutron scattering data\cite{Kearley1,Kearley2,Kearley3,Kearley4,Kearley5}
are clearly inferior to the other experimental datasets. This is hardly surprising as 
this method relies on fitting and scaling a harmonic force field which is 
obtained from DFT in small split-valence basis sets. Already for the harmonics, this introduces errors 
on the order of about 50 cm$^{-1}$. 
The error for the C-H stretches is expected to be even larger. 
It would be interesting to reanalyze their data based on the present anharmonic force fields, although this
still would not resolve the problems associated
with the C-H stretches, for which large basis set coupled cluster frequencies would be desirable.
 The experimental IR/Raman data of Daunt et al.\cite{Shurvell} and Lancaster et
al.\cite{Colthup} are very similar to those of IRM\cite{Moomaw} and are not listed in
Table \ref{tab8}. With the exception of the first $e''$ $\nu_{12}$ and the C-H stretches,
which are again underestimated, all computed values are within 11 cm$^{-1}$ of experiment
Again, the errors are similar to the other azabenzenes; including the C-H stretches, the
mean absolute error is 7.6 cm$^{-1}$ and the RMS error 11 cm$^{-1}$. Excluding the
stretches (and $\nu_{12}$), these numbers reduce to 5.1 (3.4) and 7.5 (5.0) cm$^{-1}$,
respectively.

\subsection{1,2,3-Triazine}

For 1,2,3-Triazine (Figure 1g), only one IR and Raman spectrum in KBr has been
experimentally determined\cite{Scott}: the assignment was assisted by scaled MP2/6-31G* harmonic
frequencies. The spectrum looks very similar to that of pyrimidine; even the
same Fermi resonances are present.\\
 At first glance at the spectra in Table \ref{tab9}, the agreement between theory
and experiment is disappointing. For very few of the modes is there agreement
between the spectra. However, the authors did not consider {\em any} Fermi resonances
in the lower spectrum (although they noted that one might exist, involving $\nu_{17}$), 
and simply assumed the ordering of the fundamentals to be identical to that of the MP2/6-31G*
harmonics.  This however leads to obvious misassignments.
For example, $\nu_6$ and $\nu_{18}$ 
lie almost on top of each other, while $\nu_{12} + \nu_{13}$ borrows intensity from $\nu_6$
because of the Fermi resonance, as can be seen in Table \ref{tab10}.
If the resonance were not as strong as computed --- specifically, if the unperturbed fundamental were
shifted only a few cm$^{-1}$ towards higher values, the perturbed $\nu_{12} + \nu_{13}$,
presently calculated at 1056 cm$^{-1}$, might easily correspond to the band at 
1064 cm$^{-1}$, assigned to $\nu_6$ in Ref.\cite{Scott}. The same actually applies to $\nu_{20}$,
which has a strong Fermi resonance with $\nu_{9}$ + $\nu_{13}$, that has, however, been
calculated at a lower value of 644 cm$^{-1}$. We have also reassigned the first $a_1$ C-H
stretch (mode 1), since the discrepancy between experiment and theory appears too high even 
for a C-H stretch. Needless to say, this mode also has a strong Fermi resonance with the
$2 \nu_{15}$ overtone, which has its largest component for the eigenvector with 
associated eigenvalue at
at 3100 cm$^{-1}$, again explaining the experimental value perfectly. The only other
value which is still in question would be a small peak at 1597 cm$^{-1}$, which might
correspond to an impurity; a band was found that was assigned to acetylene, a 
fragmentation product of 1,2,3-triazine. The value of 318 cm$^{-1}$, which has been
reported for $\nu_{13}$ is actually just a scaled Hartree-Fock harmonic frequency.\\
Therefore, a dramatic change in the errors is obtained by reassigning the experimental
spectrum. Including the C-H stretches in the error analysis, the mean absolute error
decreases from 17.7 to 4.7 cm$^{-1}$ and the RMS error from 27.9 to 6.1 cm$^{-1}$.
By excluding the C-H stretches, the mean absolute and RMS errors improve
from 14.6 and 22.1 cm$^{-1}$, respectively, to 4.1 and 5.2 cm$^{-1}$. Thus, 
accuracy consistent with the preceding azabenzenes can be obtained following some
reassignments in the experimental spectrum. This also corroborates our conclusion 
concerning pyridazine that neither substitution of C-H by N
nor N-N bonds seem to be responsible for the discrepancies between experimental and
computed vibrational spectra.

\subsection{1,2,4-Triazine}

Because 1,2,4-triazine (Figure 1h) is highly unstable, only one ``full'' experimental
spectrum has been measured as a liquid film\cite{Hennig}, although more recently two
high-resolution IR spectra have been recorded for four frequencies\cite{Palmer1,Palmer2}.
For the C-H stretches (Table \ref{tab11}), possibly $\nu_1$ and $\nu_2$ may have been swapped,
since they both take part in a large resonance matrix block. 
Overall, ten frequencies and
combination bands interact forming this large resonance block that include modes 1 to 3.
Hence, it is not surprising that  a correct assignment based merely on experimental data would be problematic.
With this swap, all three C-H stretches are
underestimated by the usual 10-30 cm$^{-1}$. Only $\nu_{11}$ appears questionable in view
of our results. However, $\nu_9$ has a Fermi resonance with the combination of
$\nu_{17}$ and $\nu_{21}$, and the resonance, eigenvector and eigenvalue matrices are
displayed in Table \ref{tab12}. The Fermi resonances displayed in
these tables are similar to pyrimidine, 1,2,3-triazine and 1,2,4-triazine. Assuming that
both of the Fermi resonances are slightly stronger than predicted, since $\nu_{21}$ is
underestimated by 3 cm$^{-1}$ compared to the high-resolution IR data, the new
$\nu_{17}$ + $\nu_{21}$ could well end up lower than the predicted 1143 cm$^{-1}$, 
explaining the value of 1136 cm$^{-1}$ that would then not be assigned to any mode.
The assignment of $\nu_{11}$ in the experiment would then correspond to $\nu_{10}$, in agreement
with the calculated spectrum. The assignment of $\nu_{13}$ also appears to be incorrect.\\
Of the four modes obtained by high-resolution IR, the newer data for $\nu_{12}$ 
validates our computational value of 1044 cm$^{-1}$ rather than the
1050 cm$^{-1}$ predicted by the older experiment. On the other hand, the low-resolution IR
value of 769 cm$^{-1}$ for mode 19 is closer to the hi-res data compared to our
predicted 779 cm$^{-1}$. These new experiments show the usefulness of the theoretical value,
since its resolution is apparently even comparable to that obtained by low-resolution IR data.
\\ 
 Without making any reassignments to the experimental spectrum, the mean absolute error 
from experiment is 13.6 cm$^{-1}$, and the RMS error is 19.2 cm$^{-1}$ including the
C-H stretches. This appears to be quite large in comparison to the other azabenzenes.
With the abovementioned suggested reassignments, these errors change to 7.5 cm$^{-1}$ and
10.0 cm$^{-1}$, respectively. Excluding the C-H stretches, we once again end up
with a mean absolute error of 5.1 cm$^{-1}$ and an RMS error of 5.8 cm$^{-1}$.

\subsection{1,2,4,5-Tetrazine (s- Tetrazine)}

For 1,2,4,5-Tetrazine (Figure 1i), CCSD(T) harmonic frequencies are available
in the literature\cite{Luthi}. Unfortunately, they were calculated with all
electrons correlated in a basis set (6-311G**) that is only minimal in the 
inner-shell orbitals. 
The resulting errors in  harmonic frequencies can easily exceed
20 cm$^{-1}$ for small molecules\cite{C4}, several times larger than the actual
effect of neglecting core correlation (typically less than 10 cm$^{-1}$ in HCNOF
systems\cite{core}). We have recomputed the CCSD(T)/6-311G** frequencies correlating
only valence electrons (Table \ref{tab13}). For the in-plane modes, differences
with the all-electron calculations are quite minor (6 cm$^{-1}$ or less). The out-of-plane
modes, however, are drastically affected, up to 60 cm$^{-1}$ (Table \ref{tab13}, especially the $a_{u}$ and the
second b$_{3u}$ frequencies). This adds yet another stanza to the long litany of rejoinders
in the literature (e.g.\cite{TaylorLNC1992}) against the practice of correlating inner-shell electrons in 
basis sets unsuitable for the purpose.\\
We have recalculated the harmonic frequencies at the CCSD(T) level (frozen core)
with basis sets of $spdf$ quality, specifically
Dunning's popular cc-pVTZ \cite{Dun89} and the Alml\"of-Taylor atomic natural
orbital\cite{Alm87} basis set used in Ref.\cite{benzccsdt} for benzene. The
CCSD(T)/cc-pVTZ and CCSD(T)/ANO4321 frequencies (Table \ref{tab13}) are in
excellent agreement with each other, the largest difference being 6 cm$^{-1}$ 
for the lowest frequency ($\omega_{18}$). (This is very unlike the case of
benzene\cite{benzccsdt}, where the two $b_{2g}$ modes are hypersensitive to the
basis set and differ significantly even between cc-pVTZ and ANO4321.) . The two corresponding modes in s-tetrazine display appreciable 
differences between 6-311G* and cc-pVTZ but barely between cc-pVTZ and ANO4321.)   
In contrast, the CCSD(T)/6-311G** frequencies differs significantly from
both data sets, particularly for the two $b_{2g}$ modes (39 and 33 cm$^{-1}$ w.r.t.
ANO4321), followed by the lowest $b_{2u}$ mode (a.k.a. `Kekul\'e mode', 27 cm$^{-1}$) and the
$a_u$ mode (25 cm$^{-1}$, but a relative error of 7\%). So even for a valence-only CCSD(T) calculation, the 6-311G** basis set is insufficient, and the RMS
deviation from the CCSD(T)/ANO4321 results, 16.4 cm$^{-1}$ (17.3 when excluding the C-H stretching frequencies), is comparable to that
of the much more cost-effective B97-1 calculations,  18.1 (15.1) cm$^{-1}$.

Turning now to the B97-1/TZ2P harmonic frequencies, the most notable differences 
with the CCSD(T)/ANO4321 data are for the CH stretches (underestimates of 34 cm${-1}$ for both) and the Kekul\'e mode (overestimated by 37 cm${-1}$). Possibly, self-interaction error would adversely affect\cite{Ziegler} 
the presently 
used functional's description of the potential surface along the Kekul\'e mode, and
it would be interesting how a self-interaction corrected functional would perform. Unfortunately no
implementation of such a functional is available to the authors, let alone an implementation including
analytical derivatives. Exclusive of this mode and the CH stretches, the RMS deviation between the 
CCSD(T)/ANO4321 and B97-1/TZ2P frequencies is about 11 cm$^{-1}$, compared to just 4 cm$^{-1}$ for
benzene. 

As expect, agreement between B97-1/TZ2P fundamentals and experiment is likewise compromised
(Table \ref{tab14}). We note that s-tetrazine is both the least stable and the most `inorganic' molecule
of the series, and that it has a low-lying excited state at less than 2 eV\cite{Luthi}.
If we combine the CCSD(T)/ANO4321 harmonic frequencies with the B97-1/TZ2P anharmonic
corrections, the picture brightens somewhat. Compared to the assignment of IRM\cite{Moomaw},
the RMS deviation amounts to 9.8 cm$^{-1}$ if the symmetric C--H stretch is excluded. 
The peculiarly large splitting of 76 cm$^{-1}$ between IRM's symmetric and asymmetric CH stretches
is impossible to reconcile with any of our calculations, all of which suggest a very small 
splitting on the order of just 3 cm$^{-1}$, in agreement with the earlier film IR and Raman study of
Sigworth and Pace (SP)\cite{Sigworth}. (Their band origin of 3089 cm$^{-1}$ for the symmetric CH stretch
is quite compatible with our calculations: IRM's value of 3010 cm$^{-1}$ was taken from single-vibronic-level
fluorescence spectra\cite{Innes1}. An older gas-phase study by Franks, Merer, and Innes (FMI)\cite{Innes2} found both
CH stretches at 3090 cm$^{-1}$; the symmetric one in the IR, the antisymmetric one in the Raman spectrum.)
Our calculations do not reveal a resonance interaction affecting $\nu_1$ 
that is severe enough to make a large CH symmetric/antisymmetric spllitting plausible.
We hence conclude that $\nu_1$ of IRM is erroneous.

One other IRM assignment which is difficult to reconcile with our best calculations concerns
the highest $b_{3u}$ mode (calculated: 902, IRM: 929 cm$^{-1}$)
for which the Franks, Merer, and Innes\cite{Innes2} value of 904 cm$^{-1}$
nearly perfectly matches our calculation. It would appear that even the somewhat degraded 
performance, exhibited by our model for this molecule, still ought to be quite useful for
resolving assignment issues or guiding high-resolution spectroscopic measurements.

\section{Conclusions}

We have assessed the performance of modern density functional theory (particularly,
B97-1/TZ2P) for the anharmonic vibrational spectra of a series of related medium-sized
molecules. 
The C-H stretching frequencies are consistently underestimated by about 20-30 cm$^{-1}$;
disregarding them,
the level of agreement with experiment that can be achieved is quite astonishing (on the
order of 6 cm$^{-1}$ RMS deviation for fundamentals).
Somewhat poorer accuracy is achieved for pyridazine and particularly for s-tetrazine.
For several systems (particularly 1,2,3-triazine), our calculations strongly suggest
revised assignments of the observed frequencies.
 For pyridazine, unfortunately, the experimental results are mutually contradictory and
our analysis inconclusive. For s-tetrazine, basically all predicted fundamentals agree
less well with experiment; the more `inorganic' character of the molecule
may have adversely affected the performance of our calculations.
Nevertheless, even here a RMS deviation of
about 15 cm$^{-1}$ for DFT and 10 cm$^{-1}$ for the combined CCSD(T)/DFT results is obtained.\\
Overall, B97-1/TZ2P quartic force fields combined with second-order rovibrational 
perturbation theory shows great promise for the assignment of vibrational spectra
of medium-sized organic molecules.

\section{Acknowledgments}
A. D. Boese is grateful for financial support 
from
the Feinberg Graduate School.
 This
research was supported by the Lise Meitner-Minerva Center for Computational Quantum Chemistry, of which
J. M. L. Martin is a member, and the Helen and Martin A. Kimmel Center for Molecular Design.

\section*{Supporting information}

B97-1/TZ2P quartic force fields in normal coordinates of the 
molecules studied are available for download (in SPECTRO format)
at the URL \url{http://theochem.weizmann.ac.il/web/papers/azabenzenes.html}.

\indent

\newpage
\pagestyle{empty}
\clearpage
\begin{figure}
\includegraphics{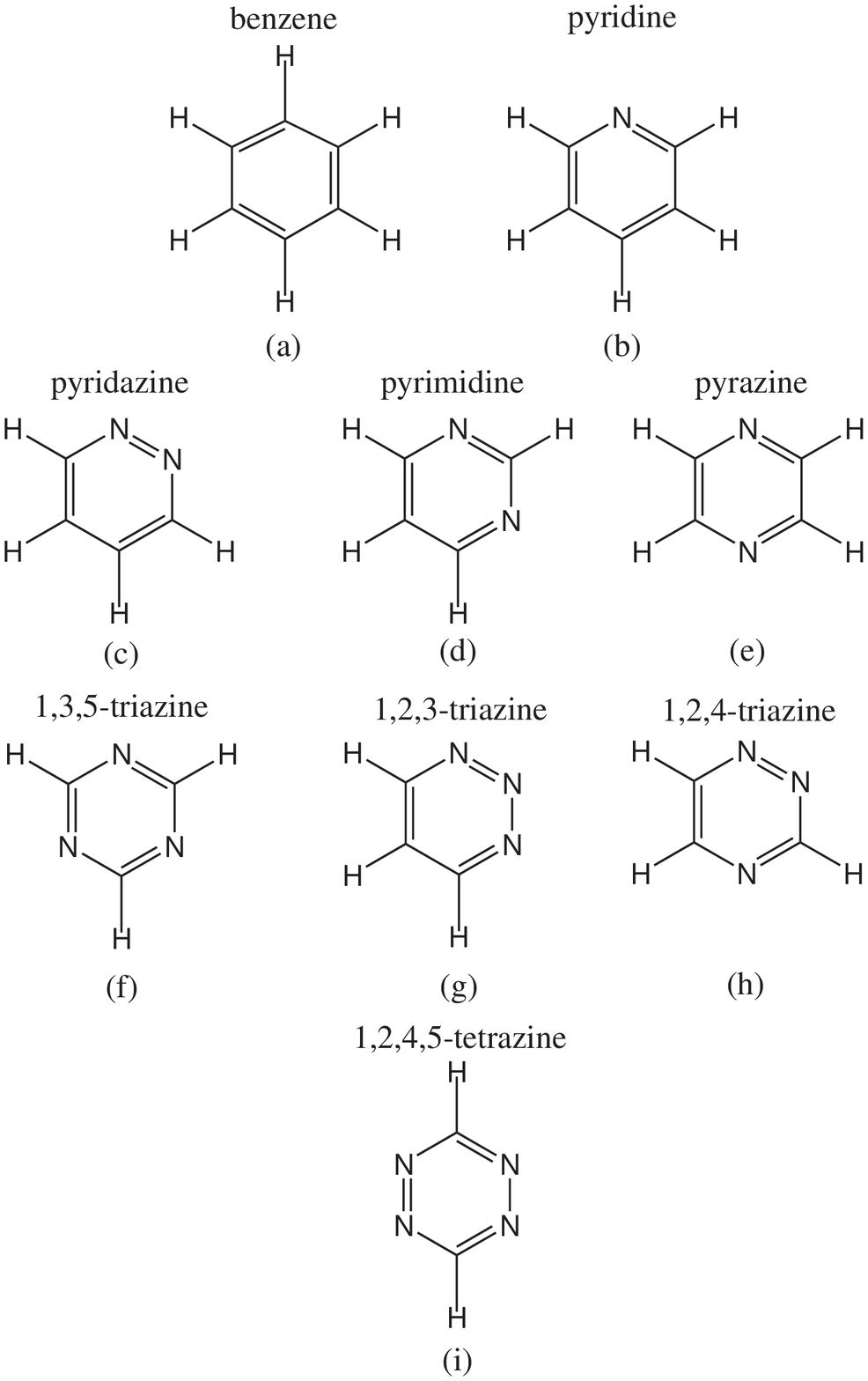}
\caption{\label{fig1}Boese et al., Journal of Physical Chemistry A}
\end{figure}

\newpage
\clearpage
\noindent
Fig. 1:\\
The molecules investigated in this study, including benzene and the azabenzene series up to 1,2,4,5-tetrazine.\\

\newpage
\clearpage
\begin{table}
\caption{ Harmonic frequencies of benzene. \label{tab1}}
\begin{tabular}{|l|l|l|l|l|l|l|}\hline\hline
         &\multicolumn{3}{c|}{Calculated}&\multicolumn{3}{c|}{Derived from Experiment}\\\hline
         & B97-1 & B3LYP\cite{anharmhandy} & CCSD(T)\cite{benzccsdt} & HMA$^a$\cite{HMA} & HW$^b$\cite{HW} & MCPTH$^d$\cite{anharmhandy} \\ \hline
\multicolumn{7}{|c|}{In-Plane}\\\hline
a$_{1g}$ & 3189  & 3192      & 3210    & 3198  & 3191 & 3218 \\\hline
         & 1008  & 1012      & 1003    & 1001  & 1008 & 1003 \\\hline
a$_{2g}$ & 1380  & 1392      & 1380    & 1378  & 1367 & 1392 \\\hline
b$_{2g}$ & 1012  & 1017      & 1009    & 1000  &  990 & 1012 \\\hline
         &  718  &  723      &  709    &  712  &  718 &  717 \\\hline
e$_{2g}$ & 3165  & 3168      & 3183    & 3182  & 3174 & 3210 \\\hline
         & 1628  & 1635      & 1637    & 1623  & 1607 & 1645 \\\hline
         & 1195  & 1201      & 1194    & 1185  & 1178 & 1197 \\\hline
         &  617  &  624      &  611    &  610  &  613 &  617 \\\hline
e$_{1g}$ &  863  &  864      &  865    &  856  &  847 &  861 \\\hline
a$_{2u}$ &  686  &  686      &  687    &  680  &  686 &  683 \\\hline
b$_{1u}$ & 3155  & 3159      & 3173    & 3173  & 3174 &      \\\hline
         & 1022  & 1031      & 1020    & 1016  & 1024 & 1030 \\\hline
\multicolumn{7}{|c|}{Out-of-Plane}\\\hline
b$_{2u}$ & 1330  & 1333      & 1326    & 1313  & 1318 & 1338 \\\hline
         & 1170  & 1178      & 1163    & 1158  & 1167 & 1163 \\\hline
e$_{2u}$ &  987  &  992      &  985    &  978  &  967 &  987 \\\hline
         &  408  &  412      &  406    &  402  & 1058 & 1056 \\\hline
e$_{1u}$ & 3180  & 3183      & 3200    & 3186  & 3181 & 3212 \\\hline
         & 1509  & 1519      & 1509    & 1503  & 1494 & 1522 \\\hline
         & 1055  & 1060      & 1056    & 1048  & 1058 & 1057 \\\hline\hline
\multicolumn{7}{c}{$^a$ $\omega_{av}$ Handy, Murray and Amos,  based on the experiments of Goodman et al.\cite{GOT}.}\\
\multicolumn{7}{c}{$^b$ Handy and Willets, experimental $\nu_i$ (Goodman et al.\cite{GOT}) - HF $(\nu_i-\omega_i)$.}\\
\multicolumn{7}{c}{$^b$ Miani, Cane, Palmieri, Trombetti, and Handy, experimental $\nu_i$ - B3LYP/TZ2P $(\nu_i-\omega_i)$.}\\
\end{tabular}
\end{table}
\newpage
\clearpage
\begin{table}
\caption{ Computed and observed fundamental frequencies of benzene. For the deviations experiment-theory, values in
parentheses are exclusive of C-H stretching frequencies. \label{tab2}}
\begin{tabular}{|l|l|l|l|l|l|l|}\hline\hline
         &      & \multicolumn{3}{c|}{Calculated}&\multicolumn{2}{c|}{Experiment}\\\hline
         & Mode & B97-1/TZ2P & B3LYP/TZ2P & CCSD(T)/B97-1 & GOT\cite{GOT} & MCP\cite{anharmhandy} \\\hline
\multicolumn{7}{|c|}{In-Plane}\\\hline
a$_{1g}$ &   1  & 3048       & 3051       & 3069 & 3074 & 3073 \\\hline
         &   2  &  992       &  995       &  987 &  993 &  993 \\\hline
a$_{2g}$ &   3  & 1348       & 1351$^*$   & 1348 & 1350 & 1350 \\\hline
b$_{2g}$ &   7  &  987       &  997       &  984 &  990 &  993 \\\hline
         &   8  &  707       &  708       &  698 &  707 &  702 \\\hline
e$_{2g}$ &  15  & 3032$^*$   & 3028       & 3050 & 3057 & 3057 \\\hline
         &  16  & 1611$^*$   & 1613$^*$   & 1620 & 1601 & 1610 \\\hline
         &  17  & 1178       & 1181       & 1177 & 1178 & 1178 \\\hline
         &  18  &  613       &  615$^*$   &  607 &  608 &  608 \\\hline
e$_{1g}$ &  11  &  841       &  846       &  843 &  847 &  847 \\\hline
a$_{2u}$ &   4  &  672       &  677       &  673 &  674 &  674 \\\hline
b$_{1u}$ &   5  & 3004$^*$   & 2988$^*$   & 3022 & 3057 & 3057 \\\hline
         &   6  & 1006       & 1015       & 1004 & 1010 & 1014 \\\hline
\multicolumn{7}{|c|}{Out-of-Plane}\\\hline
b$_{2u}$ &   9  & 1309       & 1305       & 1305 & 1309 & 1309 \\\hline
         &  10  & 1156       & 1163       & 1149 & 1150 & 1148 \\\hline
e$_{2u}$ &  19  &  964       &  972       &  962 &  967 &  968 \\\hline
         &  20  &  400       &  403       &  398 &  398 &  398 \\\hline
e$_{1u}$ &  12  & 3031$^*$   & 3023$^*$   & 3051 & 3064 & 3048 \\\hline
         &  13  & 1486       & 1484       & 1486 & 1484 & 1484 \\\hline
         &  14  & 1045       & 1038       & 1046 & 1038 & 1038 \\\hline
\multicolumn{2}{|c|}{B3LYP} &\multicolumn{3}{|c|}{RMS deviation experiment-theory}& 15.2 (4.2) & 17.0 (4.8) \\\hline
\multicolumn{2}{|c|}{B97-1} &\multicolumn{3}{|c|}{RMS deviation experiment-theory}& 12.2 (2.8) & 13.7 (4.5) \\\hline
\multicolumn{2}{|c|}{CCSD(T)/B97-1} &\multicolumn{3}{|c|}{RMS deviation experiment-theory}&  7.8 (4.5) &  7.4 (4.1) \\\hline\hline
\multicolumn{7}{c}{$^*$ Band affected by Fermi resonance.}\\
\end{tabular}
\end{table}

\newpage
\clearpage
\begin{table}
\scriptsize{
\caption{ Computed and observed fundamental frequencies of pyridine (infrared intensities in km/mol for IR
active modes are given in parentheses). All RMS deviations exclude C-H stretches. \label{tab3}}
\begin{tabular}{|l|l|l|l|l|l|l|l|l|l|l|}\hline\hline
      &      & \multicolumn{3}{c|}{Calculated}&\multicolumn{6}{c|}{Experiment}\\\hline
      &      & B3LYP/pVTZ& \multicolumn{2}{c|}{B97-1/TZ2P}&\multicolumn{2}{c|}{IR}&\multicolumn{3}{c|}{Raman}& INS \\\hline
      & Mode & Harmonic\cite{first} & Harmonic & Fundamental & Liquid & Vapor & Vapor & Liquid & Solution & Refined \\ \hline
      & Mode & Ref.\cite{first} &\multicolumn{2}{c|}{this work}& Ref.\cite{Stidham} & Ref.\cite{Colson} & Klots\cite{Klots} & Klots\cite{Klots} & Ref.\cite{Kakiuti} & Ref.\cite{Kearley1}\\ \hline
\multicolumn{11}{|c|}{In-Plane}\\\hline
a$_1$ &  1   & 3195(7.4) & 3191(4.4) & 3054$^*$   & 3070   & 3094(0.0)         & 3094  & 3090   &     & 3089 \\\hline
      &  2   & 3171(5.2) & 3168(4.0) & 3043$^*$   & 3057   & 3073(1.5$\pm$1.0) & 3067  & 3056   &     & 3075 \\\hline
      &  3   & 3147(6.8) & 3147(3.5) & 3012$^*$   & 3025   & 3030(8.5$\pm$1.0) & 3030  & 3021   &     & 3057 \\\hline
      &  4   & 1626(23.8)& 1620(23.3)& 1575$^*$   & 1581   & 1584(17.9$\pm$1.8)& 1590  & 1588   &     & 1582 \\\hline
      &  5   & 1518(3.2) & 1512(3.4) & 1483       & 1483   & 1483(4.0$\pm$0.4) & 1483  & 1482   &     & 1482 \\\hline
      &  6   & 1244(3.2) & 1240(3.5) & 1221       & 1217   & 1218(4.3$\pm$0.4) & 1218  & 1217   &     & 1209 \\\hline
      &  7   & 1096(3.0) & 1091(3.6) & 1071       & 1069   & 1072(4.5$\pm$0.5) & 1072  & 1068   &     & 1058 \\\hline
      &  8   & 1052(7.7) & 1044(6.8) & 1028       & 1030   & 1032(7.7$\pm$0.8) & 1032  & 1031   &     & 1030 \\\hline
      &  9   & 1012(5.7) & 1006(4.9) &  990       &  991   &  991(5.4$\pm$0.5) &  991  &  991   &     &  991 \\\hline
      &  10  &  617(4.7) &  610(4.1) &  604       &  603   &  601(4.4$\pm$0.4) &  601  &  603   &     &  603 \\\hline
b$_2$ &  14  & 3186(26.7)& 3183(17.3)& 3090$^*$   & 3079   & 3094(15.9$\pm$1.6)& 3087  & 3079   &     & 3034 \\\hline
      &  15  & 3144(29.9)& 3145(20.7)& 2971$^*$   & 3034   & 3042(5.1$\pm$1.5) & 3042  & 3035   &     & 3018 \\\hline
      &  16  & 1621(8.9) & 1614(8.3) & 1575       & 1574   & 1581(7.3$\pm$1.8) & 1581  & 1573   &     & 1580 \\\hline
      &  17  & 1477(27.2)& 1468(28.0)& 1442       & 1437   & 1442(31.1$\pm$3.1)& 1442  & 1438   &     & 1437 \\\hline
      &  18  & 1391(0.1) & 1384(0.1) & 1356       & 1355   & 1362(0.5$\pm$0.2) & 1355  & 1355   &     & 1355 \\\hline
      &  19  & 1283(0.07)& 1280(0.0) & 1237$^*$   & 1227   & 1227(0.0)                & 1225  & 1228   &     & 1230 \\\hline
      &  20  & 1173(2.4) & 1166(2.5) & 1155       & 1146   & 1143(3.6$\pm$0.4) & 1143  & 1147   &     & 1137 \\\hline
      &  21  & 1080(0.1) & 1073(0.01)& 1045       & 1069   & 1079(0.0)                & 1052  & 1053   &     & 1045 \\\hline
      &  22  &  670(0.2) & 664(0.3)  &  658       &  654   &  652(1.1$\pm$0.2) &  654  &  653   &     &  654 \\\hline
\multicolumn{11}{|c|}{Out-of-Plane}\\\hline
b$_1$ &  23  & 1023(0.02)& 1012(0.02) &  993      & 1007   & 1007(0.0)                &  991  &  995   & 988 & 1005 \\\hline
      &  24  &  964(0.01)&  957(0.04) &  940      &  941   &  937(0.0)                &  937  &  941   & 931 &  948 \\\hline
      &  25  &  769(5.9) &  761(9.9)  &  749      &  747   &  744(12.9$\pm$1.3)&  744  &  749   & 744 &  748 \\\hline
      &  26  &  721(63.1)&  715(69.4) &  704      &  703   &  700(67.5$\pm$6.7)&  700  &  708   & 700 &  710 \\\hline
      &  27  &  422(3.5) &  414(3.5)  &  406      &  406   &  403(7.2)                &  403  &  407   & 405 &  406 \\\hline
a$_2$ &  11  & 1011      &  1003      &  980      &  980   &  966                     &  982  &  982   & 979 &  984 \\\hline
      &  12  &  899      &   895      &  877      &  884   &  871                     &  875  &  884   & 873 &  895 \\\hline
      &  13  &  385      &   379      &  371      &  380   &  373                     &  371  &  379   & 374 &  380 \\\hline
\multicolumn{2}{|c|}{B97-1} &\multicolumn{3}{|c|}{RMS deviation experiment-theory}& 6.5 & 8.4 & 5.3 & 5.2 & & \\\hline\hline
\multicolumn{11}{c}{$^*$ Band affected by Fermi resonance.}\\
\end{tabular}}
\end{table}

\newpage
\normalsize
\clearpage
\begin{table}
\caption{ Computed and observed fundamental frequencies of pyridazine. \label{tab4}}
\begin{tabular}{|l|l|l|l|l|l|l|l|l|l|l|}\hline\hline
      &      & \multicolumn{2}{c|}{Calculated}&\multicolumn{7}{c|}{Experiment}\\\hline
      &      & \multicolumn{2}{c|}{B97-1/TZ2P}&\multicolumn{3}{c|}{IR}&\multicolumn{3}{c|}{Raman}& INS \\\hline
      & Mode &Harmonic&Fundamental& Ref.\cite{Tucci} & Ref.\cite{Ozono} & Ref.\cite{Holly} & Ref.\cite{Boggs} & Ref.\cite{Tucci} & Ref.\cite{Ozono} & Ref.\cite{Kearley2} \\ \hline
\multicolumn{11}{|c|}{In-Plane}\\\hline
a$_1$ &  1   & 3193  & 3061$^*$    & 3068   &      & 3082  & 3086  & 3064   & 3070 & 3086 \\\hline
      &  2   & 3166  & 3034$^*$    & 3056   &      & 3052  & 3071  & 3052   & 3053 & 3071 \\\hline
      &  3   & 1604  & 1558        & 1570   & 1555 & 1568  & 1570  & 1572   & 1570 & 1547 \\\hline
      &  4   & 1477  & 1446        & 1415   & 1440 & 1418  & 1444  & 1417   & 1441 & 1465 \\\hline
      &  5   & 1180  & 1145        &        & 1340 & 1337  & 1160  & 1347   & 1352 & 1188 \\\hline
      &  6   & 1171  & 1157        & 1159   & 1153 & 1154  & 1119  & 1160   & 1150 & 1177 \\\hline
      &  7   & 1088  & 1064        & 1061   & 1055 & 1061  & 1061  & 1063   & 1063 & 1075 \\\hline
      &  8   & 1006  &  987        &  963   &  960 &  969  &  968  &  964   &  963 &  998 \\\hline
      &  9   &  676  &  669        &  629   &  622 &  622  &  668  &  630   &  632 &  670 \\\hline
b$_2$ &  14  & 3180  & 3056$^*$    & 3085   &      & 3082  & 3079  & 3083   & 3080 & 3079 \\\hline
      &  15  & 3162  & 3052$^*$    & 3056   &      & 3080  & 3057  & 3052   & 3041 & 3057 \\\hline
      &  16  & 1601  & 1556        & 1563   & 1540 & 1572  & 1563  & 1566   & 1564 & 1571 \\\hline
      &  17  & 1438  & 1406        & 1446   & 1408 & 1412  & 1413  & 1450   & 1401 & 1436 \\\hline
      &  18  & 1311  & 1284        & 1283   & 1283 & 1281  & 1281  & 1283   & 1287 & 1306 \\\hline
      &  19  & 1081  & 1058        & 1131   & 1112 & 1131  & 1049  & 1129   & 1113 & 1075 \\\hline
      &  20  & 1056  & 1039        &        & 1058 & 1058  & 1027  & 1032   & 1052 & 1049 \\\hline
      &  21  &  629  &  622        &  664   &  663 &  673  &  622  &  667   &  660 &  628 \\\hline
\multicolumn{11}{|c|}{Out-of-Plane}\\\hline
b$_1$ &  22  &  981  &  962        &        &      &  985  &  987  &  842   &  986 &  998 \\\hline
      &  23  &  762  &  751        &  760   &  760 &  745  &  745  &  759   &  755 &  782 \\\hline
      &  24  &  370  &  362        &  369   &  372 &  369  &  376  &  370   &  370 &  375 \\\hline
a$_2$ &  10  & 1019  & 1002        &        &      &  963  & 1025  &  938   &  970 & 1039 \\\hline
      &  11  &  943  &  926        &        &      &  765  &  945  &  861   &  786 &  949 \\\hline
      &  12  &  775  &  763        &        &      &  785  &  729  &  753   &  775 &  793 \\\hline
      &  13  &  372  &  365        &        &      &  377  &  367  &  410   &  363 &  372 \\\hline\hline
\multicolumn{11}{c}{$^*$ Band affected by Fermi resonance.}\\
\end{tabular}
\end{table}

\newpage
\clearpage
\begin{table}
\caption{ Computed and observed fundamental frequencies of pyrimidine. All RMS deviations exclusive of C-H stretches
and $\nu_{20}$ (see text). \label{tab5}}
\begin{tabular}{|l|l|l|l|l|l|l|}\hline\hline
      &      & \multicolumn{2}{c|}{Calculated}&\multicolumn{3}{c|}{Experiment}\\\hline
      &      & \multicolumn{2}{c|}{B97-1/TZ2P}&\multicolumn{2}{c|}{IR}& INS \\\hline
      & Mode &Harmonic&Fundamental& IRM\cite{Moomaw}& Ref.\cite{Holly}& Ref.\cite{Kearley3}\\ \hline
\multicolumn{7}{|c|}{In-Plane}\\\hline
a$_1$ &  1   & 3194  & 3074$^*$    & 3074   & 3082 & 3074 \\\hline
      &  2   & 3161  & 3045$^*$    & 3052   & 3053 & 3052 \\\hline
      &  3   & 3147  & 3025$^*$    & 3038   & 3053 & 3038 \\\hline
      &  4   & 1607  & 1565        & 1570   & 1572 & 1580 \\\hline
      &  5   & 1433  & 1404        & 1398   & 1465 & 1390 \\\hline
      &  6   & 1160  & 1139        & 1147   & 1155 & 1139 \\\hline
      &  7   & 1077  & 1071$^*$    & 1065   & 1065 & 1080 \\\hline
      &  8   & 1004  &  989        &  991   &  969 &  972 \\\hline
      &  9   &  691  &  685$^*$    &  678   &  679 &  665 \\\hline
b$_2$ &  12  & 3151  & 3040$^*$    & 3086   & 3047 & 3086 \\\hline
      &  13  & 1605  & 1562        & 1568   & 1569 & 1565 \\\hline
      &  14  & 1494  & 1464        & 1466   & 1411 & 1470 \\\hline
      &  15  & 1393  & 1365        & 1370   & 1356 & 1376 \\\hline
      &  16  & 1252  & 1225        & 1225   & 1224 & 1225 \\\hline
      &  17  & 1211  & 1173        & 1159   & 1158 & 1136 \\\hline
      &  18  & 1091  & 1072        & 1071   & 1071 & 1021 \\\hline
      &  19  &  630  &  623        &  623   &  621 &  628 \\\hline
\multicolumn{7}{|c|}{Out-of-Plane}\\\hline
b$_1$ &  20  & 1026  & 1006        &  980   & 1033 & 1070 \\\hline
      &  21  &  981  &  962        &  955   &  980 &  980 \\\hline
      &  22  &  822  &  809        &  811   &  804 &  826 \\\hline
      &  23  &  736  &  723        &  721   &  719 &  722 \\\hline
      &  24  &  347  &  340        &  344   &  347 &  344 \\\hline
a$_2$ &  10  & 1001  &  982        & [927]  &  960 & 1020 \\\hline
      &  11  &  404  &  396        &  399   &  398 &  344 \\\hline
\multicolumn{2}{|c|}{B97-1} &\multicolumn{2}{|c|}{RMS deviation experiment-theory}& 5.3 & 21.2 (9.9$^a$) & \\\hline\hline
\multicolumn{7}{c}{$^*$ Band affected by Fermi resonance.}\\
\multicolumn{7}{c}{$^a$ Additionally excluding $\nu_5$ and $\nu_{14}$.}\\
\end{tabular}
\end{table}

\newpage
\clearpage
\begin{table}
\caption{Fermi resonances of pyrimidine. \label{tab6}}
\begin{tabular}{|l|l|l|}\hline
\multicolumn{3}{|c|}{Resonance matrices of pyrimidine.}\\\hline
Deperturbed & $\nu_{7}$ & $\nu_{23}$ + $\nu_{24}$\\\hline
         & 1060.0  &                    \\\hline
         &  -10.9  & 1060.1             \\\hline
\multicolumn{3}{c}{}\\\hline
Deperturbed & $\nu_{9}$ & 2 $\nu_{24}$\\\hline
         &  682.7  &                    \\\hline
         &    3.7  &  680.0             \\\hline
\multicolumn{3}{c}{}\\
\multicolumn{3}{c}{}\\
\multicolumn{3}{c}{}\\\hline
\multicolumn{3}{|c|}{Eigensolutions}\\\hline
Perturbed             & $\nu_{7}$ & $\nu_{23}$ + $\nu_{24}$\\\hline
Coefficient of      &  1049.2 &  1070.9            \\\hline
  $\nu_{7}$           & -0.7085 & -0.7057            \\\hline
$\nu_{23}$ + $\nu_{24}$ & -0.7057 & -0.7085            \\\hline
\multicolumn{3}{c}{}\\\hline
Perturbed             & $\nu_{9}$  & 2 $\nu_{24}$\\\hline
Coefficient of        &   677.4 &   685.3            \\\hline
  $\nu_{9}$           & -0.5760 &  0.8175            \\\hline
2 $\nu_{24}$ &  0.8175 & -0.5760            \\\hline
\end{tabular}
\end{table}

\newpage
\clearpage
\begin{table}
\caption{ Computed and observed fundamental frequencies of pyrazine. All RMS deviations exclusive of C-H stretches and $\nu_{21}$.\label{tab7}}
\begin{tabular}{|l|l|l|l|l|l|l|l|}\hline\hline
      &      & \multicolumn{2}{c|}{Calculated}&\multicolumn{4}{c|}{Experiment}\\\hline
      &      & \multicolumn{2}{c|}{B97-1/TZ2P}&\multicolumn{3}{c|}{IR}&INS\\\hline
        & Mode &Harmonic&Fundamental & IRM\cite{Moomaw} & Ref.\cite{Holly} & Ref.\cite{Marcos} & Ref.\cite{Kearley4}\\ \hline
\multicolumn{8}{|c|}{In-Plane}\\\hline
a$_g$   &  1   & 3168  & 3043$^*$    & 3055   & 3053 & 3054 & 3054 \\\hline
        &  2   & 1612  & 1567        & 1580   & 1579 & 1574 & 1571 \\\hline
        &  3   & 1251  & 1233        & 1233   & 1235 & 1230 & 1240 \\\hline
        &  4   & 1037  & 1020        & 1016   & 1015 & 1015 & 1015 \\\hline
        &  5   &  607  &  598        &  602   &  601 &  596 &  602 \\\hline
b$_{3g}$&  11  & 3147  & 3024$^*$    & 3040   & 3062 & 3040 & 3040 \\\hline
        &  12  & 1576  & 1533        & 1525   & 1522 & 1529 & 1529 \\\hline
        &  13  & 1372  & 1342        & 1346   & 1353 & 1343 & 1359 \\\hline
        &  14  &  715  &  706        &  704   &  698 &  701 &  698 \\\hline
b$_{1u}$&  15  & 3148  & 2990$^*$    & 3012   & 3017 & 3018 & 3018 \\\hline
        &  16  & 1513  & 1481        & 1483   & 1484 & 1482 & 1485 \\\hline
        &  17  & 1163  & 1136        & 1130   & 1135 & 1135 & 1062 \\\hline
        &  18  & 1030  & 1013        & 1018   & 1020 & 1020 & 1007 \\\hline
b$_{2u}$&  19  & 3162  & 3055$^*$    & 3069   & 3069 & 3069 & 3069 \\\hline
        &  20  & 1440  & 1413        & 1411   & 1413 & 1416 & 1427 \\\hline
        &  21  & 1214  & 1174        & 1149   & 1337 & 1146 & 1146 \\\hline
        &  22  & 1084  & 1061        & 1063   & 1063 & 1062 & 1131 \\\hline
\multicolumn{8}{|c|}{Out-of-Plane}\\\hline
a$_u$   &   6  & 1002  &  982        & [960]  &      & [997]& 1065 \\\hline
        &   7  &  347  &  338        & [350]  &      & [422]&  350 \\\hline
b$_{1g}$&   8  &  946  &  927        &  927   &  925 &  919 &  940 \\\hline
b$_{2g}$&   9  &  987  &  968        &  983   &  976 &  983 &  983 \\\hline
        &  10  &  775  &  762        &  756   &  755 &  754 &  754 \\\hline
b$_{3u}$&  23  &  804  &  791        &  785   &  785 &  786 &  818 \\\hline
        &  24  &  429  &  421        &  418   &  417 &  420 &  414 \\\hline
\multicolumn{2}{|c|}{B97-1} &\multicolumn{2}{|c|}{RMS deviation experiment-theory}& 6.9 (5.3$^a$)& 7.1 (6.7$^a$) & 6.5 (5.5$^a$)& \\\hline\hline
\multicolumn{8}{c}{$^*$ Band affected by Fermi resonance.}\\
\multicolumn{8}{c}{$^a$ Additionally excluding $\nu_9$.}\\
\end{tabular}
\end{table}
\newpage
\clearpage
\begin{table}
\caption{Computed and observed fundamental frequencies of 1,3,5-triazine. All RMS deviations exclusive of C-H stretches.\label{tab8}}
\begin{tabular}{|l|l|l|l|l|l|l|}\hline\hline
      &      & \multicolumn{2}{c|}{Calculated}&\multicolumn{3}{c|}{Experiment}\\\hline
      &      & \multicolumn{2}{c|}{B97-1/TZ2P}&\multicolumn{2}{c|}{IR,Raman}&INS\\\hline
       & Mode &Harmonic&Fundamental & Ref.\cite{Crawford}& IRM\cite{Moomaw}& Ref.\cite{Kearley5} \\ \hline
\multicolumn{7}{|c|}{In-Plane}\\ \hline
a$_1'$  &  1   & 3164  & 3025$^*$    & 3082   & 3042 & 3042 \\\hline
        &  2   & 1156  & 1137        & 1130   & 1137 & 1125 \\\hline
        &  3   & 1005  &  989        &  989   &  989 &  991 \\\hline
a$_2'$  &  4   & 1400  & 1375        &[1556]  &      & 1375 \\\hline
        &  5   & 1168  & 1124        &[1381]  &      & 1000 \\\hline
e'      &  8   & 3159  & 3034$^*$    & 3081   & 3059 & 3056 \\\hline
        &  9   & 1594  & 1545        & 1560   & 1556 & 1555 \\\hline
        &  10  & 1439  & 1405        & 1404   & 1410 & 1414 \\\hline
        &  11  & 1194  & 1167        & 1168   & 1173 & 1165 \\\hline
        &  12  &  688  &  675        &  666   &  676 &  675 \\\hline
\multicolumn{7}{|c|}{Out-of-Plane}\\\hline
a$_2''$ &   6  &  947  &  932        &        &  925 &  940 \\\hline
        &   7  &  755  &  740        &        &  737 &  732 \\\hline
e''     &  13  & 1043  & 1017        &        & 1034 & 1030 \\\hline
        &  14  &  345  &  338        &        &  339 &  333 \\\hline
\multicolumn{2}{|c|}{B97-1} &\multicolumn{2}{|c|}{RMS deviation experiment-theory}& & 7.5 (5.0$^a$)& \\\hline\hline
\multicolumn{7}{c}{$^*$ Band affected by Fermi resonance.}\\
\multicolumn{7}{c}{$^a$ Additionally excluding $\nu_{13}$.}\\
\end{tabular}
\end{table}
\newpage
\clearpage
\begin{table}
\caption{Computed and observed fundamental frequencies of 1,2,3-triazine. All RMS deviations exclusive of C-H stretches.\label{tab9}}
\begin{tabular}{|l|l|l|l|l|l|l|}\hline\hline
      &      & \multicolumn{2}{c|}{Calculated}&\multicolumn{3}{c|}{Experiment}\\\hline
      &      & \multicolumn{2}{c|}{B97-1/TZ2P}& IR & Raman & IR reassigned\\\hline
        & Mode &Harmonic&Fundamental & Ref.\cite{Scott} & Ref.\cite{Scott} & \\ \hline
\multicolumn{7}{|c|}{In-Plane}\\  \hline
a$_1$   &  1   & 3195  & 3032$^*$    & 3107   & 3110 & 3046 \\\hline
        &  2   & 3165  & 3028        &        & 3045 &      \\\hline
        &  3   & 1589  & 1548        & 1597   & 1594 & 1546 \\\hline
        &  4   & 1384  & 1347        & 1336   & 1329 & 1336 \\\hline
        &  5   & 1136  & 1122        & 1080   & 1088 & 1124 \\\hline
        &  6   & 1092  & 1081$^*$    & 1069   & 1064 & 1080 \\\hline
        &  7   & 1006  &  985        &  979   &  977 &  979 \\\hline
        &  8   &  675  &  667        &  660   &  660 &  660 \\\hline
b$_2$   &  14  & 3171  & 3042$^*$    & 3046   &      & 3046 \\\hline
        &  15  & 1584  & 1538        & 1545   & 1547 & 1545 \\\hline
        &  16  & 1438  & 1410        & 1410   &      & 1410 \\\hline
        &  17  & 1219  & 1199        & 1195   & 1198 & 1195 \\\hline
        &  18  & 1101  & 1078        & 1124   & 1127 & 1080 \\\hline
        &  19  &  974  &  929        &  935   &      &  935 \\\hline
        &  20  &  662  &  659$^*$    &  653   &      &  660 \\\hline
\multicolumn{7}{|c|}{Out-of-Plane}\\\hline
b$_1$   &  10  & 1017  &  998        &        &      &      \\\hline
        &  11  &  824  &  811        &  819   &      &  819 \\\hline
        &  12  &  783  &  769        &  769   &      &  769 \\\hline
        &  13  &  302  &  296        & [318]  &      &      \\\hline
a$_2$   &   8  &  992  &  973        &        &      &      \\\hline
        &   9  &  358  &  350        &        &  365 &      \\\hline
\multicolumn{2}{|c|}{B97-1} &\multicolumn{2}{|c|}{RMS deviation experiment-theory}& 22.1 & & 5.2 \\\hline\hline
\multicolumn{7}{c}{$^*$ Band affected by Fermi resonance.}\\
\end{tabular}
\end{table}
\newpage
\clearpage
\begin{table}
\caption{Fermi resonances of 1,2,3-triazine. \label{tab10}}
\begin{tabular}{|l|l|l|}\hline
\multicolumn{3}{|c|}{Resonance matrices of 1,2,3-triazine.}\\\hline
Deperturbed & $\nu_{6}$ & $\nu_{12}$ + $\nu_{13}$\\\hline
         & 1074.5  &                    \\\hline
         &   10.6  & 1061.4             \\\hline
\multicolumn{3}{c}{}\\\hline
Deperturbed & $\nu_{20}$&  $\nu_9$ + $\nu_{13}$  \\\hline
         &  655.1  &                    \\\hline
         &   -6.3  &  647.6             \\\hline
\multicolumn{3}{c}{}\\
\multicolumn{3}{c}{}\\
\multicolumn{3}{c}{}\\\hline
\multicolumn{3}{|c|}{Eigensolutions.}\\\hline
Perturbed            & $\nu_{6}$ & $\nu_{12}$ + $\nu_{13}$\\\hline
Coefficient of      &  1055.5 &  1080.7            \\\hline
  $\nu_{6}$           & -0.4804 & -0.8770            \\\hline
$\nu_{12}$ + $\nu_{13}$ &  0.8770 & -0.4804            \\\hline
\multicolumn{3}{c}{}\\\hline
Perturbes            & $\nu_{20}$& $\nu_{9}$ + $\nu_{13}$ \\\hline
Coefficient of      &   644.0 &   658.7            \\\hline
  $\nu_{9}$           &  0.4939 & -0.8695            \\\hline
$\nu_{9}$ + $\nu_{13}$  &  0.8175 &  0.4939            \\\hline
\end{tabular}
\end{table}
\newpage
\clearpage
\begin{table}
\caption{Computed and observed fundamental frequencies of 1,2,4-triazine. All RMS deviations exclusive of C-H stretches. \label{tab11}}
\begin{tabular}{|l|l|l|l|l|l|l|}\hline\hline
      &      & \multicolumn{2}{c|}{Calculated}&\multicolumn{2}{c|}{Experiment}\\\hline
      &      & \multicolumn{2}{c|}{B97-1/TZ2P}& IR & hires-IR \\\hline
        & Mode &Harmonic&Fundamental & Ref.\cite{Hennig} & Refs.\cite{Palmer1,Palmer2} \\ \hline
\multicolumn{6}{|c|}{In-Plane}\\  \hline
a'      &  1   & 3178  & 3051$^*$    & 3090  &\\\hline
        &  2   & 3175  & 3075$^*$    & 3060  &\\\hline
        &  3   & 3150  & 3007$^*$    & 3035  &\\\hline
        &  4   & 1596  & 1556        & 1560  &\\\hline
        &  5   & 1563  & 1524        & 1529  &\\\hline
        &  6   & 1470  & 1438        & 1435  &\\\hline
        &  7   & 1401  & 1376        & 1380  &\\\hline
        &  8   & 1314  & 1286        & 1295  &\\\hline
        &  9   & 1179  & 1171$^*$    & 1163  &\\\hline
        &  10  & 1139  & 1107        & 1136  &\\\hline
        &  11  & 1096  & 1067$^*$    & 1113  &\\\hline
        &  12  & 1063  & 1044        & 1050  & 1043.5 \\\hline
        &  13  & 1011  &  989        &  955  & \\\hline
        &  14  &  726  &  717        &  713  & \\\hline
        &  15  &  632  &  629$^*$    &       & \\\hline
\multicolumn{6}{|c|}{Out-of-Plane}\\\hline
a''     &  16  & 1014  & 1001        &       & \\\hline
        &  17  &  990  &  976        &       & \\\hline
        &  18  &  862  &  853        &  851  & \\\hline
        &  19  &  791  &  779        &  768  & 768.7 \\\hline
        &  20  &  379  &  372        &       & 367.9 \\\hline
        &  21  &  316  &  308        &       & 311.3 \\\hline
\multicolumn{2}{|c|}{B97-1} &\multicolumn{2}{|c|}{RMS deviation experiment-theory}& 16.1(5.8$^a$) &  \\\hline\hline
\multicolumn{6}{c}{$^*$ Band affected by Fermi resonance.}\\
\multicolumn{6}{c}{$^a$ Additionally excluding $\nu_{11}$; 1113 cm$^{-1}$ band reassigned to $\nu_{10}$.}\\
\end{tabular}
\end{table}
\newpage
\clearpage
\begin{table}
\caption{Fermi resonances of 1,2,4-triazine. \label{tab12}}
\begin{tabular}{|l|l|l|}\hline
\multicolumn{3}{|c|}{Resonance matrices of 1,2,4-triazine.}\\\hline
Deperturbed & $\nu_{11}$ & $\nu_{17}$ + $\nu_{21}$\\\hline
         & 1160.4  &                     \\\hline
         &  -13.7  & 1153.4                 \\\hline
\multicolumn{3}{c}{}\\\hline
Deperturbed & $\nu_{15}$&2 $\nu_{21}$  \\\hline
         &  622.8  &                    \\\hline
         &    8.4  &  617.7             \\\hline
\multicolumn{3}{c}{}\\
\multicolumn{3}{c}{}\\
\multicolumn{3}{c}{}\\\hline
\multicolumn{3}{|c|}{Eigensolutions}\\\hline
Perturbed            & $\nu_{11}$& $\nu_{17}$ + $\nu_{21}$\\\hline
Coefficient of      &  1142.7 &  1171.0            \\\hline
  $\nu_{11}$          &  0.6136 & -0.7896            \\\hline
$\nu_{17}$ + $\nu_{21}$ &  0.7896 &  0.6136            \\\hline
\multicolumn{3}{c}{}\\\hline
Perturbed            & $\nu_{15}$&2  $\nu_{21}$ \\\hline
Coefficient of      &   611.4 &   629.0            \\\hline
  $\nu_{9}$           & -0.5949 &  0.8038            \\\hline
2 $\nu_{21}$ &  0.8038 & -0.5949            \\\hline
\end{tabular}
\end{table}
\newpage
\clearpage
\begin{table}
\caption{ Harmonic frequencies of 1,2,4,5-tetrazine. \label{tab13}}
\begin{tabular}{|l|l|l|l|l|l|l|l|}\hline\hline
         &  CCSD &\multicolumn{4}{c|}{CCSD(T)}& B3LYP     & B97-1 \\
\hline
         & \multicolumn{3}{c|}{6-311G**}  & cc-pVTZ & ANO-4321 & cc-pVTZ &
TZ2P \\ \hline
         & \cite{Luthi} & Full\cite{Luthi} &
\multicolumn{3}{|c|}{Frozen-core} & \cite{first} & \\ \hline
\multicolumn{8}{|c|}{In-Plane}\\\hline
a$_{g}$  & 3253  & 3223  & 3221 & 3223 & 3226 & 3196      & 3192 \\\hline
         & 1524  & 1456  & 1452 & 1455 & 1457 & 1472      & 1473 \\\hline
         & 1059  & 1018  & 1018 & 1025 & 1025 & 1050      & 1047 \\\hline
         &  758  &  741  &  740 &  747 &  745 &  756      &  751 \\\hline
b$_{1u}$ & 3251  & 3222  & 3218 & 3222 & 3225 & 3199      & 3191 \\\hline
         & 1250  & 1212  & 1210 & 1218 & 1216 & 1232      & 1231 \\\hline
         & 1109  & 1088  & 1088 & 1096 & 1098 & 1099      & 1096 \\\hline
b$_{2u}$ & 1505  & 1471  & 1465 & 1472 & 1470 & 1479      & 1478 \\\hline
         & 1167  & 1137  & 1134 & 1138 & 1137 & 1151      & 1147 \\\hline
         &  845  &  908  &  905 &  932 &  927 &  964      &  969 \\\hline
b$_{3g}$ & 1611  & 1558  & 1552 & 1561 & 1561 & 1559      & 1560 \\\hline
         & 1345  & 1321  & 1319 & 1323 & 1323 & 1330      & 1325 \\\hline
         &  648  &  637  &  635 &  640 &  640 &  656      &  650 \\\hline
\multicolumn{8}{|c|}{Out-of-Plane}\\\hline
a$_{u}$  &  344  &  257  &  319 &  344 &  346 &  354      &  344 \\\hline
b$_{2g}$ &  985  &  947  &  957 &  998 &  996 & 1011      &  999 \\\hline
         &  803  &  772  &  783 &  816 &  816 &  837      &  829 \\\hline
b$_{3u}$ &  930  &  896  &  904 &  919 &  921 &  940      &  930 \\\hline
         &  297  &  228  &  270 &  276 &  270 &  258      &  250 \\\hline\hline
\end{tabular}
\end{table}
\newpage
\clearpage
\begin{table}
\caption{ Computed and observed fundamental frequencies of 1,2,4,5-Tetrazine. For RMS deviations, values in parentheses are exclusive of C-H stretching frequencies. \label{tab14}}
\begin{tabular}{|l|l|l|l|l|l|l|l|}\hline\hline
         &      &\multicolumn{2}{|c|}{Calculated} &\multicolumn{4}{|c|}{Experiment}\\\hline
         & Mode & B97-1      &CCSD(T)/B97-1& IRM\cite{Moomaw} & Ref.\cite{Sverdlov} & SP\cite{Sigworth} & FMI\cite{Innes2} \\\hline
\multicolumn{8}{|c|}{In-Plane}\\\hline
a$_{g}$  &   1  & 3073       & 3107       & 3010       & 3040 & 3089 & 3090 \\\hline
         &   2  & 1424       & 1408       & 1415       & 1489 & 1417$^c$ & 1418$^c$ \\\hline
         &   3  & 1024       & 1000       & 1009       &  990 & 1015 & 1017 \\\hline
         &   4  &  741       &  735       &  736       &  737 &  734 &  736 \\\hline
b$_{1u}$ &  11  & 3070       & 3104       & 3086       & 3070 & 3086 & 3090 \\\hline
         &  12  & 1204       & 1189       & 1204       & 1200 & 1204 & 1200 \\\hline
         &  13  & 1079       & 1081       & 1093       & 1106 & 1109$^d$ & 1103$^d$ \\\hline
b$_{2u}$ &  14  & 1445       & 1437       & 1448       & 1434 & 1448 & 1440 \\\hline
         &  15  & 1112$^*$   & 1102       & 1108       & 1187 & 1151 &      \\\hline
         &  16  &  927       &  885       &  883       & 1085 &  893$^d$ &  881$^d$ \\ \hline
b$_{3g}$ &   8  & 1511       & 1511       & 1525       & 1543 & 1523$^c$ & 1521$^c$ \\\hline
         &   9  & 1294       & 1293       & 1290       & 1278 & 1302 & 1303 \\\hline
         &  10  &  645       &  638       &  640       &  679 &  649 &  651 \\\hline
\multicolumn{8}{|c|}{Out-of-Plane}\\\hline
a$_{u}$  &   5  &  336       &  338       &  335       &  319 &  335 &  335 \\ \hline
b$_{2g}$ &   6  &  977       &  974       &  994       &  925 & 1015 &      \\ \hline
         &   7  &  814       &  801       &  801       &  775 &  799 &  800 \\ \hline
b$_{3u}$ &  17  &  911       &  902       &  929       &  890 &  929 &  904 \\\hline
         &  18  &  244       &  261       &  254       &  340 &  254 &  254 \\\hline
\multicolumn{4}{|c|}{RMS dev. from $\nu_i$(B97-1/TZ2P)}& 15.2$^a$(15.1,14.8$^b$) & 54.8 (57.5) & 19.0 (19.4) &16.0 (15.6)  \\
\multicolumn{4}{|c|}{$\omega_i$(CCSD(T)/ANO4321) + $\nu_i-\omega_i$(B97-1/TZ2P)} & 12.2$^a$(11.3,9.4$^b$) & 63.5 (64.5) & 20.1 (20.3) &11.0 (10.2)  \\\hline\hline
\multicolumn{7}{c}{$^*$ Band affected by Fermi resonance.}\\
\multicolumn{7}{c}{$^a$ Reassigning Mode 1 to 3090 cm$^{-1}$.}\\
\multicolumn{7}{c}{$^b$ Excluding also Mode 17.}\\
\multicolumn{7}{c}{$^c$ These assignments swapped based on IRM and our
calculations.}\\
\multicolumn{7}{c}{$^d$ These assignments swapped based on IRM and our
calculations.}\\
\end{tabular}
\end{table}

\end{document}